\documentclass{aa}  
\usepackage{graphicx}
\usepackage{txfonts}
\begin{document}
   \title{Recovery of p-modes in the combined 2004-2005 MOST\thanks{MOST is a Canadian Space Agency mission, operated jointly by Dynacon, Inc., and the Universities of Toronto and British Columbia, with assistance from the University of Vienna.} observations of Procyon}

   \author{S. V. Marchenko \inst{1}}

   \offprints{S. V. Marchenko}

   \institute{$^1$ Department of Physics and Astronomy, Western Kentucky
University, 1906 College Heights Blvd., 11077, Bowling Green, KY 42101-1077;  \email{sergey.marchenko@wku.edu}
             }

   \date{Received ...; accepted ...}

% \abstract{}{}{}{}{} 
% 5 {} token are mandatory
 
  \abstract
   {}
   {Procyon A, a bright  F5 IV-V Sun-like star, is  justifiably regarded as a prime 
   asteroseismological target. This star was repeatedly observed by MOST, a specialized microsatellite providing long-term, non-interrupted broadband photometry of bright targets. So far, the widely anticipated p modes eluded direct photometric detection, though  numerous independent approaches hinted for the presence of signals in the 
   f$\sim0.5-1.5$\,mHz range.}
   {Implementation of an alternative approach in data processing, as well as combination of the MOST data from 2004 and 2005 (264\,189 measurements in total) helps to reduce the instrumental noise affecting 
   previous reductions, bringing the $3\sigma$ detection limit down to $\sim$5.5\,part-per-million
   in the $f=0.8-1.2$\,mHz range.}
   {This enables to cross-identifiy 16 p-mode frequencies (though not their degrees) which were previously detected via high-precision radial velocity measurements, and provides an estimate of the large spacing,  $\delta\nu =0.0540$\,mHz at $f\sim1$\,mHz. The relatively
   low average amplitude of the detected modes, $a=5.8\pm0.6$\,ppm, closely matches the
   amplitudes inferred from the ground-based spectroscopy and upper limits projected from WIRE photometry. This also explains why such low-amplitude signals 
   eluded the direct-detection approach which exclusively relied  on the MOST 2004 (or 2005) data processed by a 
   standard pipeline.}
   {}


   \maketitle
%
%________________________________________________________________

\section{Introduction}
Procyon A (F5 IV-V) has served as a prime target in many asteroseismological campaigns 
(see the reviews by Bedding \& Kjeldsen, 2006,2007, and references therein).
%owing 
%to its brightness and well-established fundamental parameters (e.g., \cite{prov}), meant to be verified 
%during such campaigns.
While high-precision measurements of radial velocities revealed the  presence of
variability caused by solar-like oscillations (\cite{ma99}) and provided a preliminary 
identification of p-mode frequencies (\cite{egge}; \cite{mart}; 
 \cite{lecc}), the photometric studies met either limited success
(\cite{brun}) or outright null-detection (\cite{matt}; \cite{gun}). The null-detection result 
was hotly debated ever since (\cite{bedd};  \cite{regu}). Attempting to resolve the controversy, we 
applied an alternative data-processing approach to the abundant photometric data
acquired by MOST in 2004-2005.   
   
%__________________________________________________________________

\section{Observations and data processing}
\begin{figure*}[!ht]
\centering
\includegraphics[width=1.0\linewidth]{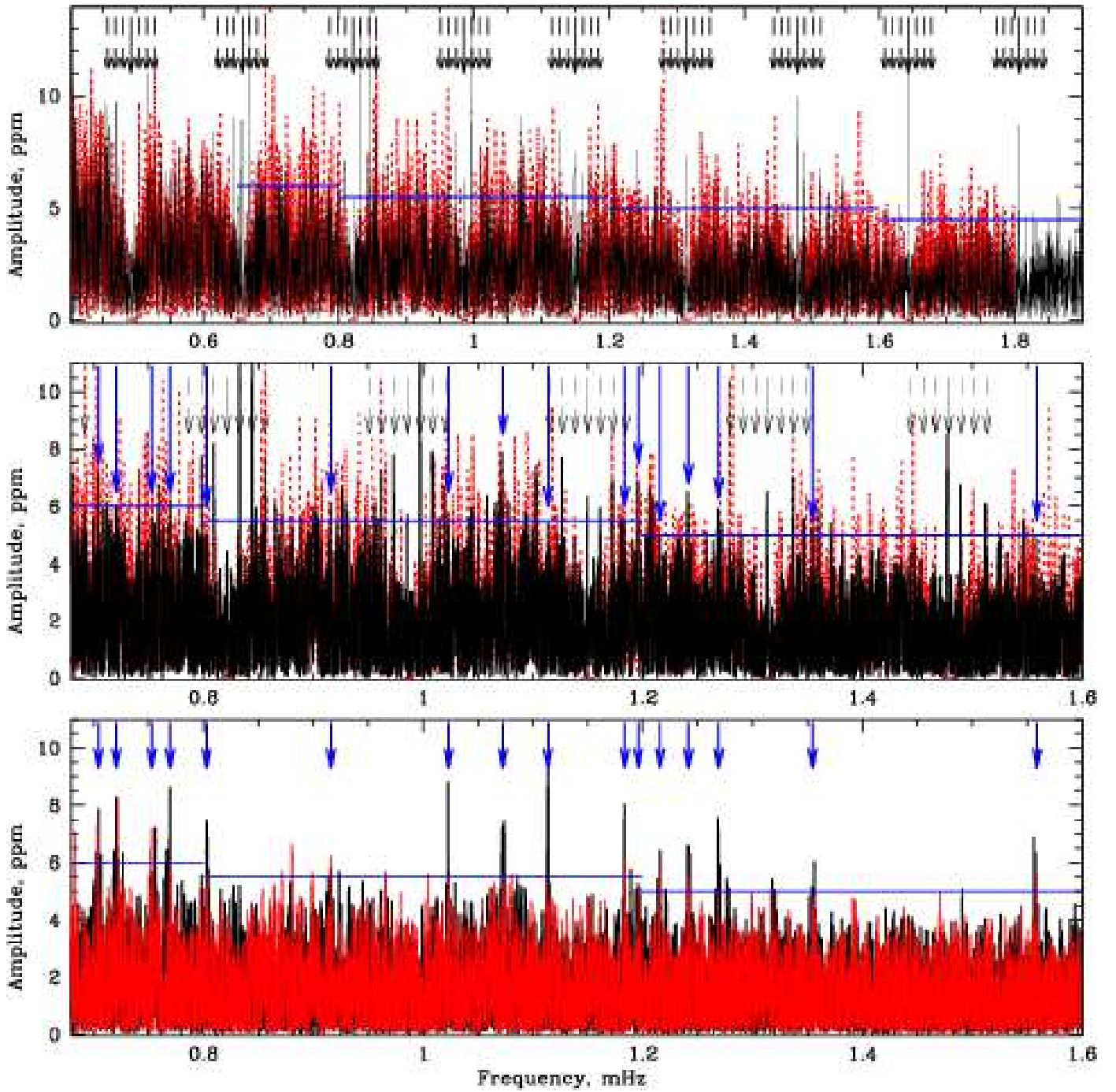}
%\vskip 4.5cm
\caption{The upper section 
shows the new spectrum (black lines) derived from the stand-alone 2004 data and  the previously published (red lines) spectrum derived from the same data. The orbital harmonics and their n*1/d sidelobes are marked by vertical solid-line and dashed-line  arrows, respectively.
Middle section: the Fourier amplitude spectrum calculated from the combined 2004-2005 data set processed with the alternative
algorithm (black lines), 
compared to the published (\cite{matt}: red lines) spectrum as a reference. Blue arrows mark the identified 
p-modes.  Bottom section: an artificial spectrum calculated from the shuffled MOST 
2004+2005 data with added stochastically-excited sinusoidal signals. Their frequencies 
are taken from Table 1. For simplicity, all the signals are assigned equal amplitudes a=10 ppm  and lifetimes L=7 days
(red lines), and  a=15 ppm, L=3 days (black lines). Blue arrows mark the identified 
p-modes. In all sections the 
horizontal blue lines denote the 3$\sigma$ detection limits.}  \label{fig1}
 \end{figure*}

The MOST satellite (\cite{walk}) is a 15-cm telescope acquiring high-precision, non-stop CCD photometric 
data on bright objects through a broad-band, 350-700 nm, optical filter. MOST observed Procyon A in 2004 (32-day non-stop observations, 231\,524 0.9 s exposures: see the results discussed 
in \cite{matt}), 2005 (taken over 17 contiguous days with the general setup of the 
experiment following the 2004 campaign: \cite{gun}) and 2007 (data 
are being processed: J. Matthews, 2007, priv. comm.). We took the publicly available 
raw data from 2004 and 2005 and processed them using an alternative algorithm
(see the description in \cite{aert}). 
While generally conforming to the approach of  the standard processing 
pipeline applied to the 2004-2005 data, 
the alternative algorithm treats the highly variable stray light, the main source of instrumental noise, in a more flexible manner. In addition to the iterative subtraction of scattered light, the algorithm 
accounted for less substantial sources of instrumental noise, such as sensitivity drift caused by 
slow variations of the CCD temperature. The alternative approach resulted in a lower point-to-point scatter,
$s=$440 part-per-million (ppm), vs.  $s=$500 ppm ( \cite{matt}) (cf. the upper section of Fig 1). This should be compared to the projected level of the Poisson-dominated instrumental noise $s_{in}=$260 ppm. The 
alternative approach also reduces the average (noise-dominated) amplitudes and scatter in the frequency spectrum: e.g., for f=1.05-1.25 mHz one obtains ${\bar {a}}(f)=2.8\pm1.7$ ppm in the 
older reduction  vs. ${\bar {a}}(f)=2.6\pm1.4$ for  the new algorithm (Fig.~\ref{fig1}). The modified
approach, when efficiently combining the two available data streams (SDS1 and SDS2: see the definition in \cite{aert}),  proves to be even more effective with the 2005 data where the better planning of experiment 
allowed more straightforward suppression of the stray-light component, providing $s\approx$375 ppm, 
which only moderately exceeds the expected level of $s_{in}=$250 ppm. Predictably, the lower noise level brings down average amplitudes in the frequency domain: $\bar a(f)=1.4\pm1.6$ ppm in the f=1.05-1.25 mHz range, twice lower than in the data from 2004. On the other hand, the standard approach (\cite{gun}), relying uniquely on the SDS2 channel, provides
a noise level comparable to the 2004 set. 

 During the processing of the 2004 data we kept the time-binning approach similar to the 
original scheme (\cite{matt}); this provided a valuable point-to-point comparison between the two different approaches and retained 214\,279 flux measurements. For the 2005 data we slightly 
changed the approach, by time-binning the SDS1 observations  centered on the less frequent SDS2 measurements. On average, taking into account the higher rejection rates,
the SDS1 data were  $\sim 5.5$ times more abundant in the 2005 set.  This time-binning
approach resulted in 49\,910 flux measurements. 

\section{Results}
Our first step is to compare the frequency spectra calculated 
by the two data-processing algorithms. As anticipated, the alternative approach, while 
providing a frequency spectrum of an overall similar appearance, reduces  the level of instrumental noise  across the spectrum, save a few rare exceptions related to the 
harmonics of the orbital period (black arrows in Fig.~\ref{fig1}).
 
The next step aims at an optimal combination of the data. A host of potential problems stems 
from: (a) the different durations of the 2004 and 2005 runs; (b) different planning of the experiments
resulting in different levels of stray light; (c) presumably unstable nature of the signals. 
In the time domain we experimented with different lengths of segments derived from the original time series, varying them from 3 days to 2 weeks and producing a weighted combination of the frequency spectra derived from the segments.
This allowes us to see that  the typical lifetime of signals does not exceed $\sim 1$ week. However,
the segment-based approach comes at high price, raising in the region of interest, 
f=0.5-2.0 mHz, the detectability limits to an unacceptably high level, $a>10$ ppm.
Hence, we finally relied on the simplest and most straightforward approach, by combining the unweighted 2004 and 2005 
data into a single set (Fig.~\ref{fig1}). Obviously, this results in  a further improvement of sensitivity:
considering the  3-sigma level counted from the average amplitude of the noise-dominated signal, one is able  to detect coherent oscillations with $a\ge 6.0$ ppm (f=0.7-0.8 mHz), $a\ge 5.5$ ppm (f=0.8-1.2 mHz), $a\ge 5.0$ ppm (f=1.2-1.6 mHz), $a\ge 4.5$ ppm (f$>$1.6 mHz), to be compared to 
 $a\ge 7.0$ ppm (f=0.5-2.0 mHz:  \cite{matt}) and $a\ge 10-11$ ppm (\cite{gun}).  Beyond the overall  improvement of sensitivity, such 
 straightforward combination of the 2004 and 2005 data should substantially increase the chance of  a positive detection which depends on the presumably 
 short  lifetimes of the coherent signals (p-modes) in Procyon A (\cite{lecc}) and the stochastic 
 nature of the mode-excitation events (\cite{kj95}). 

\begin{figure}
\centering
%\plotfiddle{fig2.eps}{2.9in}{0}{100}{100}{-300}{-285}
%\special{psfile=fig2.eps
%hoffset=35 voffset=-55 hscale=50 vscale=50 angle=90}
\includegraphics[width=0.8\linewidth]{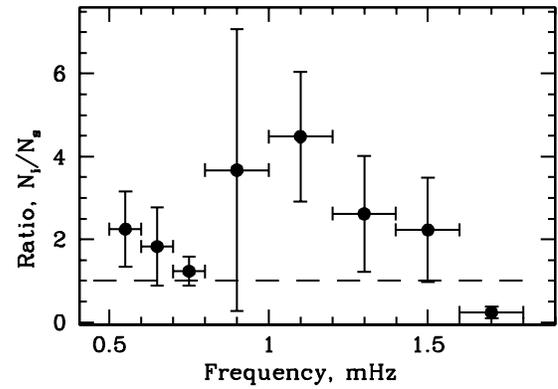}
\caption{The numbers ($N_i$) of significant (at $3\sigma$ level) peaks in the 2004+2005 AS normalized 
by the corresponding numbers ($N_s$) derived from artificial data sets. The vertical bars show the $\pm1\sigma$ errors of the ratios. The horizontal dashed line provides a reference level of 1
(the case when all peaks are generated by noise).}
 \label{fig2}
 \end{figure}

As a first test of presence of any periodic variations, we derived a Fourier amplitude spectrum (AS) of the combined 2004+2005 data in the f=0.5-2.0 mHz domain specifically targeted in 
numerous observing campaigns. Then we calculated the number of peaks exceeding the $3\sigma$
detectability levels (see above).  We also produced 10 samples via random shuffling of the 
real data and calculated the incidence of significant signals in the artificial ASa. 
We normalized the real data by the outcomes of simulations and show them in Fig.~\ref{fig2}. There is a surplus of $3\sigma$ detections around 
f$\sim$0.8-1.4 mHz, which closely matches the spread and location of the power 
excess determined from numerous independent spectroscopic runs (see \cite{lecc} and 
references therein).
    
\begin{figure}[!ht]
\centering
\includegraphics[width=0.8\linewidth]{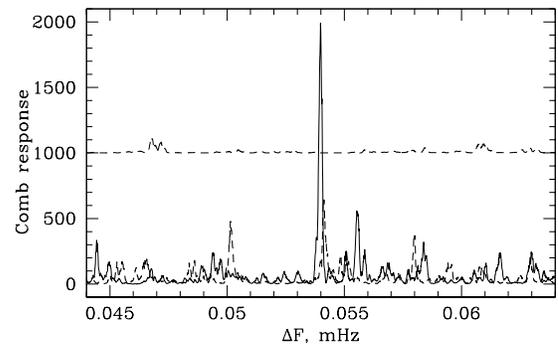}
\caption{The comb-response functions obtained for the central frequencies f=1.0726 mHz (full line), f=0.9158 mHz 
(dashed line) and f=1.6424 mHz 
(dashed line, offset for clarity).
The tallest peak corresponds to a large separation $\Delta f=0.0540\pm0.0001$\,mHz, with the error 
derived from the FWHM of the peak.}
 \label{fig3}
 \end{figure}

Further testing brings even more encouraging results. We select a prominent  feature at f=1.072 mHz (Table~\ref{table1}
and  Fig.~\ref{fig1}) and search for a comb-like pattern
(\cite{kj95et}) in the immediate vicinity of the peak. Scanning the adjacent region with $\delta f=0.0001$
mHz steps, we find a maximum response at f=1.0726\,mHz  (Fig.~\ref{fig3}), related to the large 
frequency separation of $\Delta f=0.0540\pm0.0001$\,mHz, in good agreement with   $\Delta f=0.0536-0.0559$\,mHz  consistently provided by all previous spectroscopic campaigns  (\cite{egge}; \cite{mart}; \cite{lecc}). 
The clear presence of a regularly-spaced signal in the MOST-2004  data was noted on multiple 
occasions: $\Delta f=0.0545$\,mHz (\cite{regu}),  $\Delta f\sim0.0548$\,mHz (J.Matthews, priv. comm. 2005), 
also by the referee of the article. Can the regular spacing 
of signals be caused by an unaccounted component of the stray light contributing to the $n*\nu_{orb}/3$ orbital harmonics with 0.0548\,mHz separation? Apparently not, as the newly derived $\Delta f$ is more than sufficiently distanced from the instrumental component.  Moreover, we find only a very weak trace of the instrumental signal in the comb-generated response at  f=1.0726 mHz and its complete lack at  f=1.6424 mHz (Fig.~\ref{fig3}: note that f=1.6424 mHz matches an orbital harmonic).

 The presence of a regular `comb' of periodic signals around f=1.072 mHz prompted the final step of the 
 analysis. We compiled a list of frequencies presumably related to p-modes (\cite{egge};  \cite{mart};
  \cite{lecc}) and searched for corresponding significant peaks in our 2004+2005 AS. We limited our search to the immediate 
 surroundings of the published frequencies, i.e., to the  $\pm  \delta f ^< _\sim 0.002$ mHz intervals, where $\delta f$ is an uncertainty related to frequency resolution
 in the AS derived on $\sim$1 week-long runs. Note that a formal 
 frequency resolution in the combined 2004+2005 AS reaches $3*10^{-5}$ mHz.   We recorded the maxima which were very close to or above the corresponding $3\sigma$ levels for the given frequencies. 
 Facing the rapidly rising level of instrumental+intrinsic (granulation?) noise, 
 we disregarded all positive detections with frequencies below f=0.7 mHz. 
 We also eliminated all positive detections (4 cases) which match high harmonics 
 of the orbital period and their 1/d side lobes. We present the results in Table~\ref{table1}, 
 where in parentheses  we retain some prominent signals despite their relatively large 
 deviations from the published frequencies. For a reference, we also provide the 
amplitudes of signals in the separate subsets from 2004 and 2005, even if they do not exceed detection limits. 
We compare our positive 2004+2005 detections with signals 
 exceeding (or being reasonably close to) the detectability limits in the previously published MOST-2004 data
 (\cite{matt}),
 finding  $\sim 55\%$ of positive matches, both in frequency and amplitude. On average, one may expect $2.4\pm1.2$ ($\sim 20$ in total) noise-generated signals exceeding the imposed $3\sigma$ threshold 
 in an 0.1 mHz-wide bin placed in the region of interest, f=0.7-1.6 mHz. Hence, the chance for an 
 accidental match between a noise-generated peak and a pre-determined signal  
 registered in the previous spectroscopic campaigns (\cite{egge};\cite{mart}) is $\leq 45\%$. I.e., at least 50\% of the frequencies provided in Table~\ref{table1} should be geniune.  The 16 signals identified as 
 p-modes,
 plus 20 noise-generated peaks do not account for all positive detections, 46 signals in the 
 f=0.7-1.6 mHz range. The origin of the 10 remaining peaks is unclear.

 \begin{table}
\caption{Detected p-modes}             
\label{table1}      
\centering                                       
\begin{tabular}{l  l  l l l l}           
\hline\hline                         
f [mHz] &   & \multicolumn{1}{c}{Amplitude,}  &    \multicolumn{1}{c}{$a$}    &     \multicolumn{1}{c}{[ppm] } &    References \\     
             &  2004+2005 & 2004 & 2005  & $2004^1$ &                \\
\hline                                    
0.7041 &  7.1  & 7.4  &  4.7    &  8.0 & 3 \\      
0.7206 &  6.0  & 6.0  &  5.6    & --& 2,4  \\    
0.7525 &  (5.6)  & 6.2  &  5.3    &  (6.6) & 3 \\    
(0.7692) & 6.2 & 6.3  &  4.0    & 7.5 & 3 \\       
0.8027 &  (5.0)  & 5.7  &   4.0   &-- &  2,3 \\        
0.9158 &  6.1  & 6.6  &  5.3    & 9.0 & 2,4 \\     
1.0232 &  5.8  & 6.0  &  3.6    & -- & 2,3,4 \\    
(1.0726)& 6.2  & 9.2  &  8.1    & 7.6 &  3 \\      
1.1136 &  5.4  & 5.7  &  3.9    &-- & 4  \\    
1.1837 &  5.6  & 6.5  &   2.1    &--  & 2,3 \\      
1.1959 &  6.9  & 7.8  & 3.5    &(6.2) & 2,4\\     
1.2158 &   5.0 & 5.4  &    4.2    & -- & 3 \\        
1.2415 &  6.5  & 7.6  &  5.2    &(5.2) & 3 \\      
1.2692 &  5.9  & 6.1  &  3.3    &(5.7) & 2,3,4 \\  
1.3547 &  5.2  & 5.9  &  3.3    & (6.0)&  3\\      
1.5590 &  4.9  & 4.9  &   4.4    & --  & 2,4\\      
\hline                                              
\end{tabular}
\begin{list}{}{}
\item $^1$ \cite{matt}
\item $^2$ \cite{egge}
\item $^3$ \cite{mart}
\item $^4$ \cite{lecc}; raw data and fits.
\end{list}
\end{table}

The average amplitude of the recovered signals, $a=5.8\pm0.6$\,ppm, closely matches the
$a=7.3\pm1.1$\,ppm level inferred from the ground-based spectroscopy (\cite{lecc})  and $a=8.5\pm2$\,ppm upper 
limit from WIRE photometry (\cite{brun}). Such low-amplitude signals were inevitably swamped by instrumental noise in the 
previous reduction of the stand-alone data from 2004 and 2005. The unusually low level of the 
detected signals may be related to the anticipated short life-time of the modes, $2.0\pm0.4$\,d
(\cite{lecc}). Indeed, one may re-shuffle the original 2004-2005 MOST data and add 
16 sinusoidal signals with the frequencies matching the identified p-modes from Table 1. For simplicity, we assign to the signals  the same amplitude $a_{art}$ and lifetime $L$, let the phases of the signals vary randomly, then calculate the resulting ASa  and 
show them in Fig. 1. In both shown cases, $a_{art}$=10 ppm, $L$=7 days, and $a_{art}$=15 ppm, $L$=3 days,  one may detect up to
$\sim80-90\%$ of the artificial signals. The $a_{art}=10\pm $ppm case provides the 
average amplitude $<a_{art}>=5.0\pm0.8$ ppm, and $a_{art}$=15 ppm case results in $<a_{art}>=7.1\pm1.2$ ppm and a slightly 
better detection rate, thus proving the starting assumption about the short lifetime 
of the modes. On the other hand, the relative low signal/noise ratio of the data and, presumably, 
interaction of the stochastically excited modes affects
the derived frequencies of artificial signals ($f_{art}$): $<|f_{obs}-f_{art}|>
\approx 9*10^{-5}$ mHz, to be compared to the formal frequency resolution $\delta f=3*10^{-5}$ mHz in the combined 2004+2005 data.. 
 
\begin{acknowledgements}
The author is grateful to the MOST team for making the Procyon data 
available through the Canadian Astronomy Data Centre (CADC; operated 
by the Hertzberg Institute of Astrophysics, national Research Council of Canada). 
Numerous suggestions of the anonymous referee helped to improve presentation 
of the main results.
\end{acknowledgements}

\end{document}